# A simple scheme to realize the Rice-Mele model in acoustic system

Tianzhi Xia[1,†], Xiying Fan[2,†], Qi Chen[1], Yuanlei Zhang[1], and Zhe Li[1,†]

[1]College of Physics and Electronic Engineering, Qujing Normal University, Qujing 655011, China

[2] Department of Physics, Hubei University, Wuhan 430062, China

The Rice-Mele (RM) model, as a paradigmatic extension of the Su-Schrieffer-Heeger (SSH) chain, plays a pivotal role in understanding topological phases and quantized adiabatic transport in one-dimensional systems. Its realization in acoustic systems, however, has been hindered by the need for simultaneous precise modulation of on-site potentials and couplings. In this work, we demonstrate a method to linearly tune on-site potentials and couplings, thus realizing an acoustic Rice-Mele model. During parameter evolution, the system exhibits a Thouless pump, with the acoustic field distribution adiabatically shifting from the left edge through the bulk to the right edge—fully consistent with tight-binding model predictions. Moreover, the strategy of leveraging geometric parameters to linearly and precisely control on-site potentials and couplings is highly effective and universal for designing acoustic metamaterials, and it can be extended to other classical wave systems.

**Keywords:** Rice-Mele model; Thouless pump; acoustic metamaterials

## 1. Introduction

The Rice-Mele (RM) model is a one-dimensional tight-binding model with one electronic orbital per site and two sites per unit cell, featuring alternating on-site potential and alternating nearest-neighbor hopping [1,2]. It provides a fundamental platform for exploring multiple physical phenomena, such as Thouless pumps [3–14], nonlinear phenomena [15], Floquet-engineering [10,16-19], non-hermitian effects [9], quantum entanglement [20,21], and topological solitons [2,22]. Specifically, for Thouless pumps, the RM model and another model, namely the Aubry-André-Harper (AAH) model, together constitute two fundamental paradigms [23]. However, in acoustic systems, most experimentally realized Thouless pumps to date have utilized the AAH model [24–27]. This stems from the fact that implementing the AAH model requires modulating only a single parameter—either the on-site potential or the coupling strength, which is readily achievable in acoustic systems. In contrast, realizing the RM model necessitates the simultaneous modulation of both the on-site potential and the coupling strength, varying as sine and cosine functions, respectively.

† Corresponding author. E-mail: tianzhixia@whu.edu.cn; fxy@hubu.edu.cn; zheli@mail.qjnu.edu.cn.

Previously in acoustic systems, not only Thouless pumps but also other topological effects were predominantly realized by individually modulating either the coupling strength [28–39] or the on-site potential [27,40]. This limitation originated from the conventional approach of tuning the on-site potential through resonant cavity height adjustments: Modifying the cavity height displaces the coupling tube's junction at the cavity wall, inducing mutual interference in on-site potentials and hopping amplitudes during concurrent modulation. In this work, we introduce an ingenious on-site potential modulation scheme that effectively decouples its control from coupling modulation. This breakthrough enables the realization of the acoustic RM model and a one-dimensional Thouless pump. Our dual-parameter control method exhibits universality and can serve as a foundation for implementing most tight-binding models.

## 2. Linear regulation of on-site potential and coupling in the acoustic system.

In acoustic resonator systems for realizing topological effects, the $p$-mode [Fig. 1(a)] is typically employed for coupling, with connecting tubes attached to its high-amplitude regions. The eigenfrequency $u$ of the $p$-mode is determined by both the cavity height H and sound speed $c$, and the expression is $u = c/2\text{H}$. As shown in Fig. 1(b), we designed a cavity with height $\text{H} = 32.8\ \text{mm}$ and side length $d = 10\ \text{mm}$. This geometry ensures sufficient frequency separation between $p_z$ and $p_x/p_y$ modes to prevent intermodal interference during parameter tuning. Crucially, we discovered that introducing rigid boundaries (via drilled holes) in low-amplitude regions enables effective eigenfrequency modulation. Through full-wave simulations by the commercial finite-element software (COMSOL Multiphysics), we created a hole with square cross-sectional area $S$ along the resonator's central axis. With sound speed fixed at 343 m/s, the unperturbed resonance occurs at 5.23 kHz. Figure 1(c) shows the linear decrease of resonance frequency with increasing hole area $S$ (scatter points). The fitted eigenfrequency relationship $u = 5230 - 35.26S$ Hz (red solid line) quantitatively describes this geometric tuning mechanism.

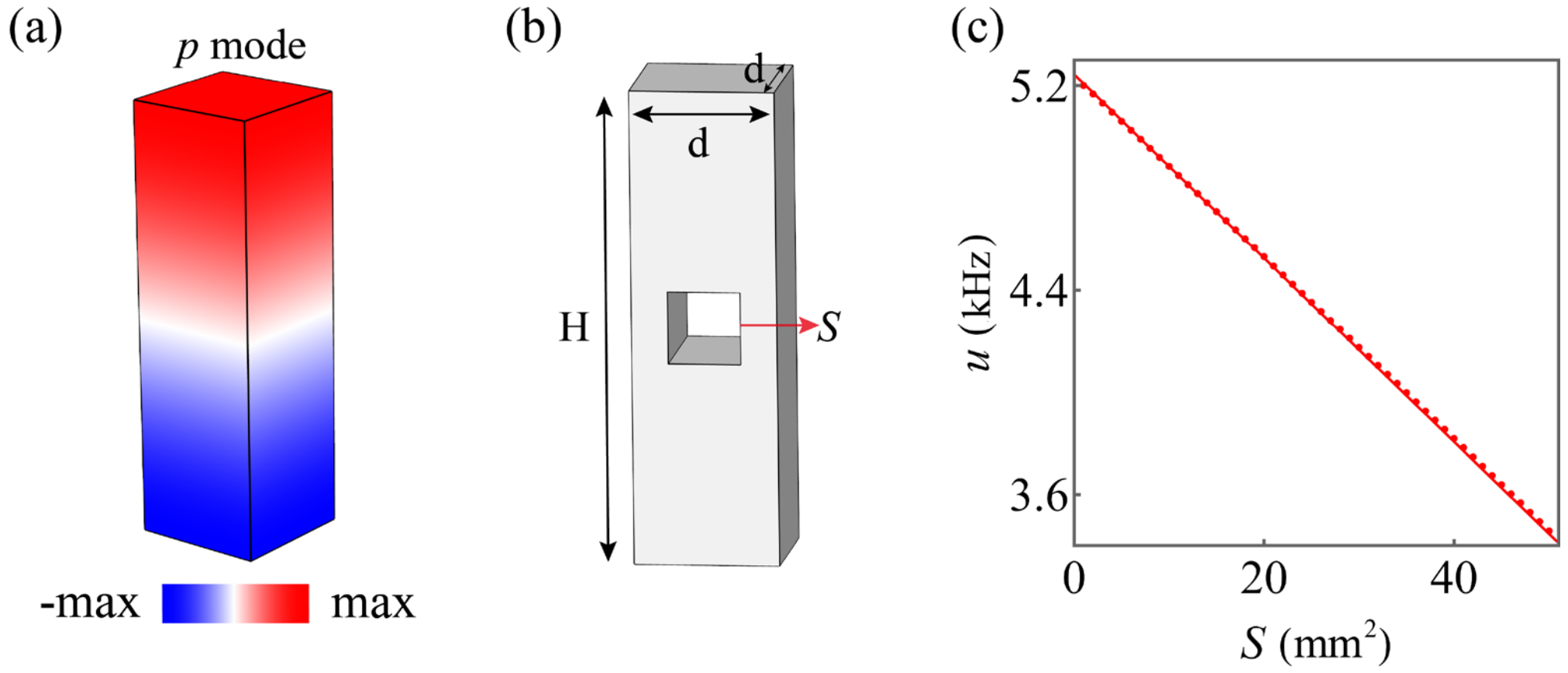

FIG. 1. Linear regulation of on-site potential. (a) Schematic diagram of the $p_z$-mode in acoustic resonator. (b) In the weak amplitude region of the $p$-mode in the resonant cavity, square holes are dug to adjust the on-site potential. (c) The on-site potential decreases linearly with the increase of the cross-sectional area of the dug holes.

In the cavity-tube model of an acoustic system, not only can the on-site potential be finely tuned in a linear manner, but the real-valued coupling can also be precisely regulated by the length and cross-sectional area of the coupling tube. As shown in Fig. 2(a), using the resonant cavity without holes mentioned above, we connect the high-amplitude regions of the $p$-mode. To ensure the chiral symmetry of the energy band, the length of the coupling tube needs to be set as a half-integer multiple of the cavity length. Specifically, with an in-phase configuration, a tube length of $(0.5+2n)\mathrm{H}$ results in negative coupling, while a tube length of $(1.5+2n)\mathrm{H}$ leads to positive coupling, where $n$ is a natural number. If the connection is made with an out-of-phase configuration in the high-amplitude regions, the sign of the corresponding coupling is exactly reversed. As depicted in Fig. 2(b), we study the relationship between the absolute value of the effective coupling and the cross-sectional area of the coupling tube when the length of the coupling tube is $0.5\mathrm{H}$. The blue scatter points represent the coupling values corresponding to specific cross-sectional areas calculated by the full-wave simulations. It can be observed that they show a linear relationship, and the fitted functional equation is $|t| = 27.992 \cdot \sigma$ (Hz) (blue solid line).

A brief explanation of why the length of the coupling tube is set as a half-integer multiple of the cavity height: Keeping the size of the resonant cavity consistent with that in the main text, we choose the cross-sectional area of the coupling tube $\sigma = 3.4\ \mathrm{mm}^2$. After fixing the cross-sectional area of the coupling tube, we change the length of the coupling tube to vary the frequency. As shown in Fig. 2(c), the dipole-mode frequency of the basic single-resonant cavity is $5.23$ kHz (cyan line). The two-cavity system can be fitted by a two-level Hamiltonian $H = \begin{pmatrix} f_0 & t \\ t & f_0 \end{pmatrix}$ [31]. Thus, the two frequencies of the system are $f_{\pm} = f_0 \pm t$, and the sound fields corresponding to the eigenstates $(1,1)^T/\sqrt{2}$ and $(1,-1)^T/\sqrt{2}$ are odd-symmetric and even-symmetric, respectively. When the frequency corresponding to the odd-symmetric state is higher than that of the even-symmetric state, it is positive coupling; otherwise, it is negative coupling. The frequency splitting between the odd and even modes as a function of the tube length is shown by the scatter points in Fig. 2 (c), where the tube length $l = \lambda H$. Taking 5.23 kHz (cyan line) as the reference, at the half-integer multiples of $\lambda$ (grey dashed lines), the split frequencies maintain chirality, that is, they are symmetric about the cyan line. Moreover, for half-integer $\lambda = 0.5 + 2n, n \in N$, it corresponds to negative coupling, and $\lambda = 1.5 + 2n$ corresponds to positive coupling. To simply represent the approximate situation of the sound field corresponding to the odd and even modes, figure 2(d) shows the pressure distributions when $\lambda = 0.5$ and $\lambda = 1.5$.

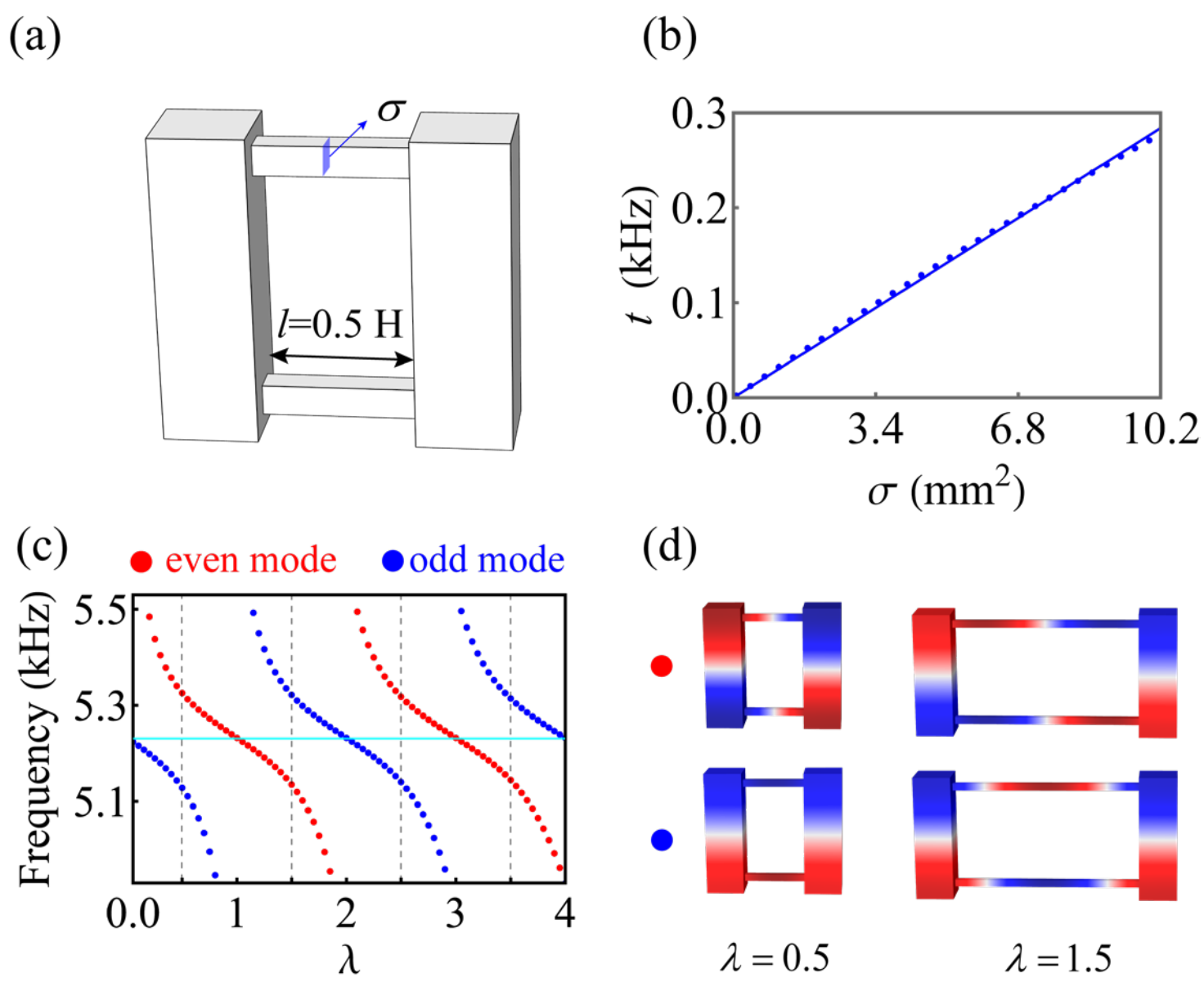

FIG. 2. Relationship between the effective coupling and the cross-sectional area of the coupling tube. (a) Schematic diagram of the cavity-tube structure. The length $l$ of the coupling tube is 0.5H. The connection in the figure is in-phase connection, corresponding to negative coupling, and the cross-sectional area of the coupling tube is $\sigma$. (b) The relationship between the magnitude of effective coupling and the cross-sectional area $\sigma$ in the case of (a). (c) Evolution of the odd-symmetric mode and the even-symmetric mode as the length of the coupling tube varies. The length of the coupling tube is $l = \lambda$ H. (b) Field patterns of the odd-symmetric mode and the even-

symmetric mode when ($\lambda = 0.5$) and ($\lambda = 1.5$).

## 3. Acoustic realization of RM model.

Given that both the on-site potential and coupling in the acoustic system can be precisely tuned, we implement the RM model using acoustic resonators. First, we introduce the RM model employed, which is derived from the Su-Schrieffer-Heeger (SSH) model [41,42,43] by introducing an additional parameter. As shown in Fig. 3(a), a single unit cell of the model contains two non-equivalent atoms with on-site potentials $u_\phi$ and $-u_\phi$ respectively. In addition, the inter-cell coupling and intra-cell coupling are $\omega_\phi$ and $v_\phi$ respectively. The on-site potential and intra-cell coupling evolve periodically with the adiabatic parameter $\phi$, that is

$$u_\phi = -\sin\phi\,, \tag{1}$$

$$v_\phi = -1 - 0.7\cos\phi\,, \tag{2}$$

$$\omega_\phi = -1. \tag{3}$$

We regard $\phi$ as a momentum dimension and calculate its bulk bands in the synthesized two-dimensional momentum space. As shown in Fig. 3(b), the bulk bands have a complete bandgap. To determine whether the system is in a topological phase capable of hosting the Thouless pump, we calculate its Chern number. Using the method of discretizing to find the Berry curvature [44], we calculate the energy band below the bandgap. The distribution of the calculated Berry curvature in the Brillouin zone is shown in Fig. 3(c). We integrate the Berry curvature over the entire first Brillouin zone and obtain the Chern number $C = -1$. This indicates that this insulating phase is topological and can host the conventional Thouless pump we aim to construct.

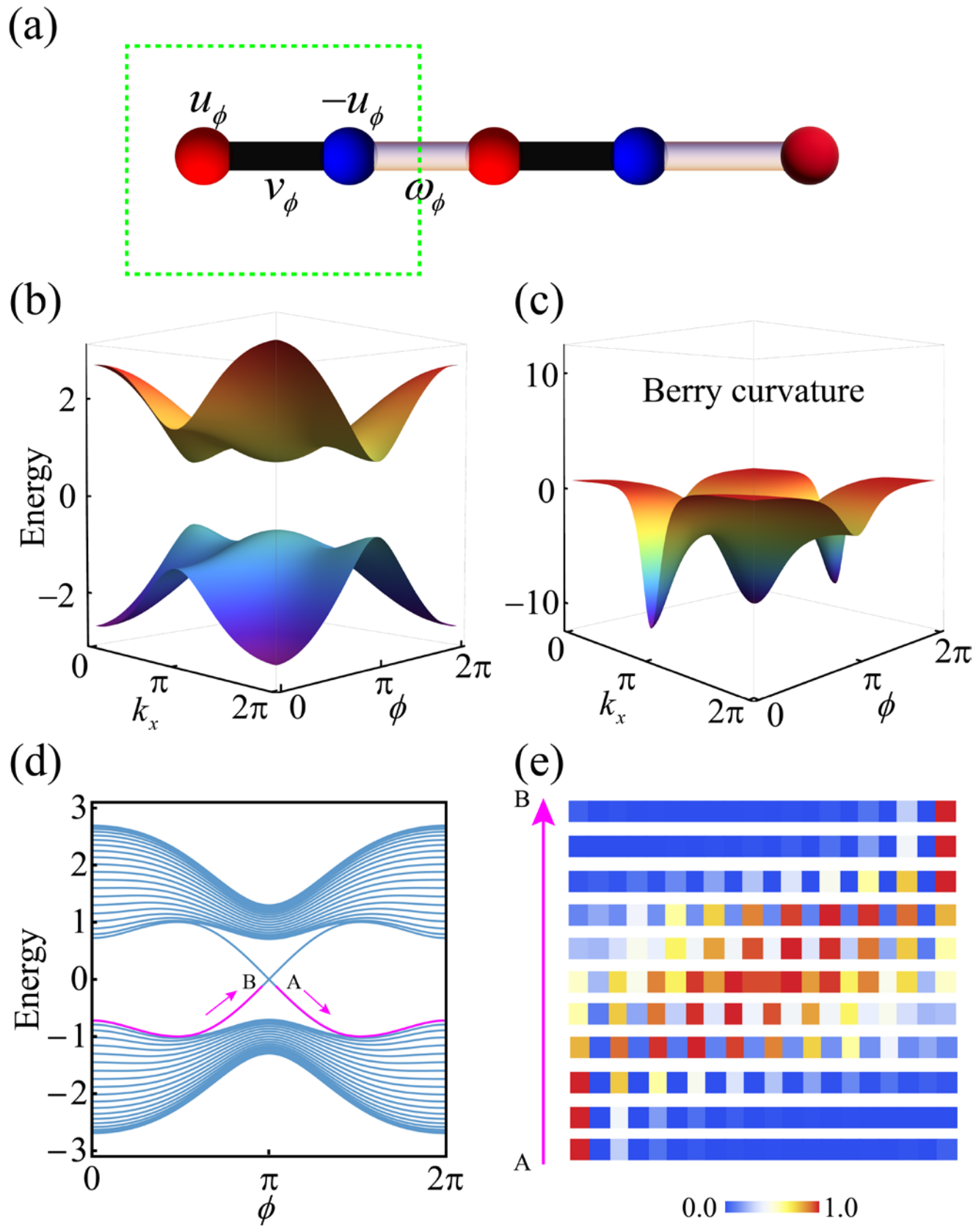

FIG. 3. The RM Model. (a) Schematic diagram of the model. (b) The synthesized two-dimensional momentum-space bulk band structure. We determine whether it is a trivial insulating phase or a topological insulating phase by calculating the Chern number of the bandgap. (c) The Berry curvature calculated using the energy band below the bandgap. The integration of Berry curvature over the entire Brillouin zone is $-1$. Therefore, the Chern number $-1$ of the bandgap characterizes a conventional first-order topological pump. (c) Energy spectrum of the finite-size RM chain. (d) As the parameters change adiabatically along a given path (pink), the system undergoes a pumping process from the left edge to the bulk state and then to the right edge. (Under a parameter path symmetric with respect to the Fermi level, the pumping process is exactly the opposite, that is, from the right edge to the bulk and then to the left edge.)

Consider the energy spectrum of the open-boundary RM chain that varies periodically with the parameter $\phi$, as shown in Fig. 3(d). Within one cycle, an edge state at the right side traverses the bandgap from the valence band to the conduction band. This indicates that in the valence band, one particle is pumped from left to right in each cycle. Selecting the adiabatic path below the Fermi level in the figure, when the parameter changes from A $(\phi = 1.05\pi)$ to B $(\phi = 0.95\pi)$, the field strength of the eigenstate corresponding to the marked energy (pink) is shown in Fig. 3(e). With the adiabatic evolution of the parameter, the eigenstate is pumped from the left edge through the bulk to the right edge.

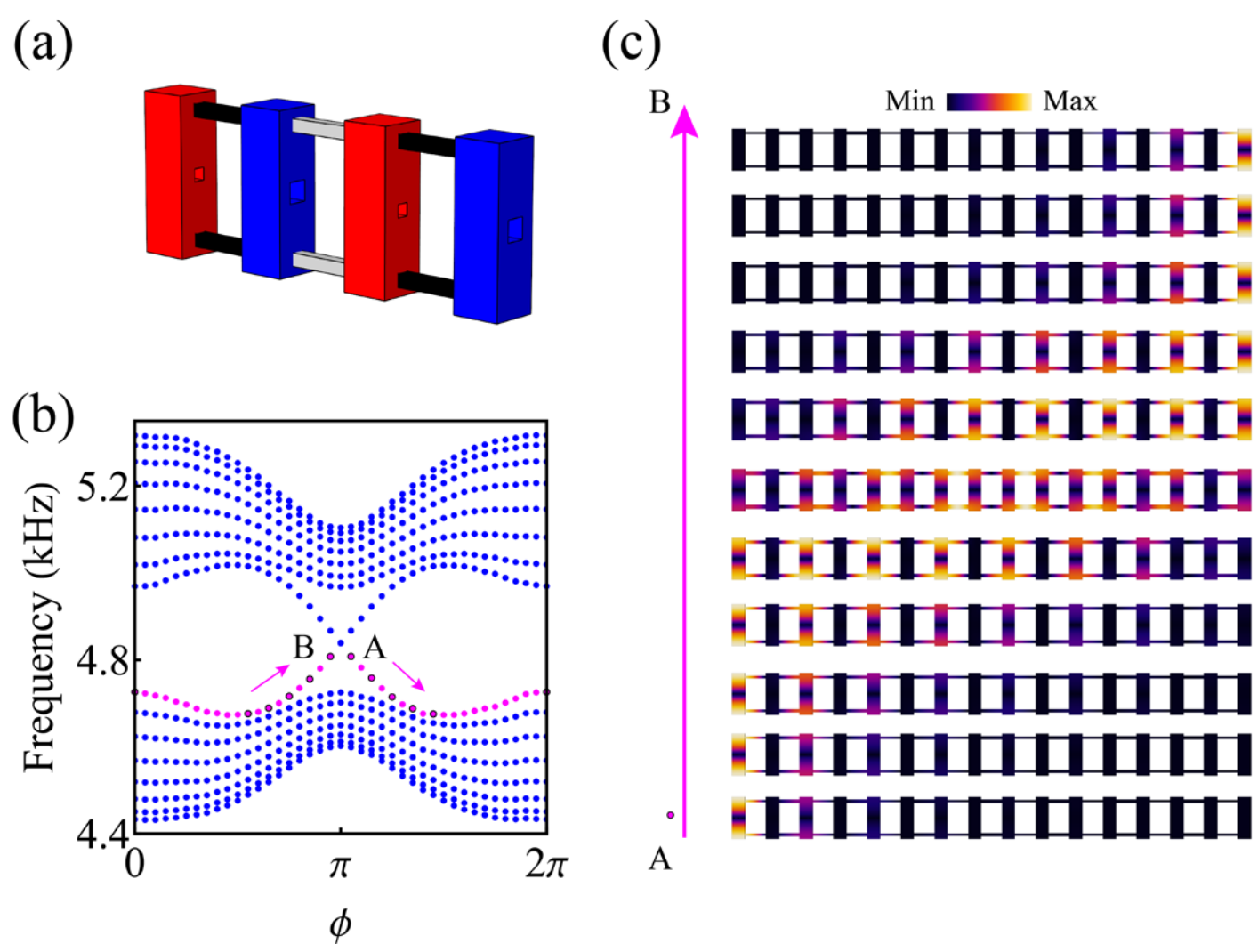

FIG. 4. Acoustic Simulation of RM model**.** (a) Structural schematic of the acoustic model. (b) Energy spectrum of the finite structure obtained from acoustic simulation. (c) Evolution of the absolute sound pressure field along a parameter path.

Based on the relationships between the on-site potential and the cross-sectional area of the dug hole, as well as the coupling and the cross-sectional area of the coupling tube, which were fitted in the previous sections, we proceed to the acoustic implementation of the acoustic RM model. As shown in Fig. 4(a), for the two resonators within a unit cell, the cross-sectional areas of the square holes are $S_1 = 12.35 - 5.56 \sin \phi$ ($\mathrm{mm}^2$) (red) and $S_2 = 12.35 + 5.56 \sin \phi$ ($\mathrm{mm}^2$) (blue). Regarding the coupling tubes connecting the resonators, the cross-sectional area of each intra-cell tube is $\sigma = 7.05 + 4.33 \cos \phi$ ($\mathrm{mm}^2$) (black), and the cross-sectional area of the inter-cell coupling tube is $\sigma' = 7.05$ ($\mathrm{mm}^2$) (gray). We use COMSOL Multiphysics to simulate the open-boundary energy spectrum of the acoustic RM model. For each chain, the number of unit cells ($n = 8$). The obtained energy spectrum is shown in Fig. 4(b), and it is in good agreement with the results of the tight-binding model. Selecting the same pumping path as that of the tight-binding model (pink scatter points, and the scatter points with black edges are selected to display the absolute sound-pressure field), and observing the sound-pressure field diagrams under discrete parameters, as shown in Fig. 4(c), when the parameter varies from A ($\phi = 1.05\pi$) to B ($\phi = 0.95\pi$), the first-order acoustic Thouless pump from the left edge state through the bulk state to the right edge state can be observed, which is completely consistent with the tight-binding prediction.

## 4. Conclusion and perspectives.

In this study, we have successfully realized the acoustic RM model and overcome the challenge of simultaneously modulating on-site potential and coupling with high precision. By introducing an innovative geometric design, specifically, drilling square holes in low-amplitude regions of resonant cavities to linearly tune on-site potentials and adjusting coupling tube cross-sectional areas to control inter-resonator couplings, we achieved the independent and linear parameter regulation in acoustic system. This decoupling strategy enabled the faithful implementation of the RM Hamiltonian, as confirmed through full-wave COMSOL simulations and tight-binding model comparisons.

Furthermore, during adiabatic parameter evolution, the system exhibited a quantized Thouless pump, with the acoustic field distribution shifting from the left edge through the bulk to the right edge, aligning perfectly with theoretical predictions. This observation validates the topological nature of the pump (Chern number of -1) and demonstrates the reliability of our approach. The universality of this dual-parameter control method extends beyond acoustics, offering a scalable framework for designing advanced metamaterials in photonic, mechanical, and other classical wave systems. By enabling precise emulation of fundamental quantum phenomena in classical wave systems, this work paves the way for experimental explorations of nonlinear effects [15] and Floquet engineering [10,16-19] in diverse platforms. Future research could leverage this methodology to investigate multi-dimensional topological phase, potentially advancing applications in wave guiding and quantum simulation. In addition, the acoustic RM model can be used to realize returning Thouless pump [45,46], and we can introduce non-Hermiticity into the acoustic RM model to achieve various novel effect [9,47-50].

**Data availability statement**

The data that support the findings of this Letter are not publicly available upon publication because it is not technically feasible and/or the cost of preparing, depositing, and hosting the data would be prohibitive within the terms of this research project. The data are available from the authors upon reasonable request.

**Acknowledgements**. This project was supported by the National Natural Science Foundation of China (Grants Nos. 12374415) and the Chutian Scholars Program in Hubei Province.

**Reference:**

[1] Rice M J and Mele E J 1982 *Phys. Rev. Lett.* **49** 145

[2] Allen R E J, Gibbons H V, Sherlock A M, Stanfield H R M, and McCann E 2022 *Phys. Rev. B* **106** 165409

[3] Nakajima S, Tomita T, Taie S, Ichinose T, Ozawa H, Wang L, Troyer M, and Takahashi Y 2016 *Nat. Phys.* **12** 296

[4] Cerjan A, Wang M, Huang S, Chen K P, and Rechtsman M C 2020 *Light Sci. Appl.* **9** 178

[5] Grinberg I H, Lin M, Harris C, Benalcazar W A, Peterson C W, Hughes T L, and Bahl G 2020 *Nat. Commun.* **11** 974

[6] Song W, You O, Sun J, Wu S, Chen C, Huang C, Qiu K, Zhu S, Zhang S, and Li T 2024 *Sci. Adv.* **10** eadn5028

[7] Danieli C, Brosco V, Pilozzi L, and Citro R 2025 *AVS Quantum Sci.* **7** 022001

[8] Bas P R and Aligia A A 2025 *Phys. Rev. B* **111** 045128

[9] Mandal I 2025 Phys. Rev. A **111** 032213

[10] Minguzzi J, Zhu Z, Sandholzer K, Walter A S, Viebahn K, and Esslinger T 2022 *Phys. Rev. Lett.* **129** 053201

[11] Dreon D, Baumgärtner A, Li X, Hertlein S, Esslinger T, and Donner T 2022 *Nature* **608** 494

[12] Citro R and Aidelsburger M 2023 *Nat. Rev. Phys.* **5** 87

[13] Cooper N R, Dalibard J, and Spielman I 2019 *Rev. Mod. Phys.* **91** 015005

[14] Zhang Y, Gu Y, Li P, Hu J, and Jiang 2022 *Phys. Rev. X* **12** 041013

[15] Bai C and Liang 2025 Phys. Rev. A **111** 042201

[16] Avdoshkin A, Mitscherling J, and Moore J 2025 *Phys. Rev. Lett*. **135** 066901

[17] Bouhon A, Timmel A, and Slager R J 2023 *arXiv*:2303.02180

[18] Jankowski W J, Morris A S, Bouhon A, Ünal F N, and Slager R J 2025 *Phys. Rev. B* **111** L081103

[19] Jankowski W J and Slager R J 2024 *Phys. Rev. Lett.* **133** 186601

[20] Lima L 2020 *Phys. Stat. Mech. Its Appl.* **545** 123800

[21] Wieder B J, et al. 2021 *Nat. Rev. Mater.* **7** 196

[22] Przysiężna A, Dutta O, and Zakrzewski 2015 *New J. Phys.* **17** 013018

[23] Ni X, Yves S, Krasnok A, and Alù 2023 *Chem. Rev.* **123** 7585

[24] Rosa M I N, Pal R K, Arruda J R F, and Ruzzene M 2019 *Phys. Rev. Lett.* **123** 034301

[25] Chen H, Zhang H, Wu Q, Huang Y, Nguyen H, Prodan E, Zhou X, and Huang 2021 *Nat. Commun.* **12** 5028

[26] Chen Z, Chen Z, Li Z, Liang B, Ma G, Lu Y, and Cheng J 2022 *New J. Phys.* **24** 013004

[27] Chen Z G, Zhu W, Tan Y, Wang L, and Ma G 2021 *Phys. Rev. X* **11** 011016

[28] Ni X, Weiner M, Alù A, and Khanikaev A 2019 *Nat. Mater.* **18** 113

[29] Xue H, Yang Y, Gao F, Chong Y, and Zhang 2019 *Nat. Mater.* **18** 108

[30] Zhang X, Xie B Y, Wang H F, Xu X, Tian Y, Jiang J H, Lu M H, and Chen Y 2019 *Nat. Commun.* **10** 5331

[31] Qi Y, Qiu C, Xiao M, He H, Ke M, and Liu Z 2020 *Phys. Rev. Lett.* **124** 206601
[32] Wu H, He H, Ye L, Lu J, Ke M, Deng W, and Liu 2025 *Adv. Mater.* **37** 2500757
[33] Yin S, Ye L, He H, Huang X, Ke M, Deng W, Lu J, and Liu 2024 *Sci. Bull.* **69** 1660
[34] Lai P, Liu Y, Pu Z, Tang Y, Liu H, Deng W, Cheng H, Liu Z, and Chen S 2025 S*ci. China Phy. Mecha. Astron.* https://doi.org/10.1007/s11433-025-2823-3
[35] Weiner M, Ni X, Li M, Alù A, and Khanikaev A B 2020 *Sci. Adv.* **6** eaay4166
[36] Xue H, Ge Y, Sun H X, Wang Q, Jia D, Guan Y J, Yuan S Q, Chong Y, and Zhang 2020 *Nat. Commun.* **11** 2442
[37] Zhang Z, Hu B, Liu F, Cheng Y, Liu X, and Christensen 2020 *Phys. Rev. B* **101** 220102
[38] Xia T, Li Y, Zhang Q, Fan X, Xiao M, and Qiu 2023 *Phys. Rev. B* **108** 125125
[39] Zheng L Y and Christensen J 2021 *Phys. Rev. Lett.* **127** 156401
[40] Ni X, Chen K, Weiner M, Apigo D J, Prodan C, Alù A, Prodan E, and Khanikaev A B 2019 *Commun. Phys.* **2** 55
[41] Zen X L, La W X, Wei Y W, and M Y Q 2024 *Chin. Phys. B* **33** 030310
[42] Li G Q, Wang B H, Tang J Y, Peng P, and Dong L W 2023 *Chin. Phys. B* **32** 077102
[43] Bao X X, Guo G F, and Tan L 2023 *Chin. Phys. B* **32** 020301
[44] Asbóth J K, Oroszlány L, and Pályi A, 2016 *A Short Course on Topological Insulators* (Cham: Springer International Publishing)
[45] Cheng Z, Yue S, Long Y, Xie W, Yu Z, Teo H T, Zhao Y X, Xue H, and Zhang B 2025 *Nat. Commun.* **16** 9669
[46] Mo Q, Liang S, Lan X, Zhu J, and Zhang S 2025 *Phys. Rev. Lett.* **135** 206603
[47] Gao H, Xue H, Gu Z, Liu T, Zhu J, and Zhang B 2021 *Nat. Commun.* **12** 1888
[48] Zhang L, et al. 2021 *Nat. Commun.* **12** 6297
[49] Wu J, Hu Y, He Z, Deng K, Huang X, Ke M, Deng W, Lu J, and Liu Z 2025 *Phys. Rev. Lett.* **134** 176601
[50] Hu Y, Wu J, Ye P, Deng W, Lu J, Huang X, Wang Z, Ke M, and Liu Z 2025 *Phys. Rev. Lett.* **134** 116606